%% file: iopconfser-AMAZE.tex
\documentclass{iopconfser}
\usepackage{orcidlink}
\usepackage{amsmath,amssymb}
\usepackage[numbers,sort&compress]{natbib}

\usepackage{acronym}
\newcommand{\chib}{\chi_{\mathrm{b}}}
\newcommand{\alphaCE}{\alpha_\mathrm{CE}}

\usepackage{etoolbox}
\newtoggle{checklength}
\toggletrue{checklength} 

\begin{document}
\acrodef{GW}[GW]{gravitational-wave}
\acrodef{BBH}[BBH]{binary black hole}
\acrodef{AMAZE}[AMA$\mathcal{Z}$E]{Astrophysical Model Analysis and Evidence Evaluation}
\acrodef{KDE}[KDE]{kernel density estimate}
\acrodef{KL}[KL]{Kullback--Leibler}
\acrodef{STFC}[STFC]{Science and Technology Facilities Council}

\input{result_macros_extra}

\title{Exploring the astrophysical origins of binary black holes using normalising flows}

\author{
S~Colloms$^{1}$\orcidlink{0009-0009-9828-3646}, 
C~P~L~Berry$^{1,2}$\orcidlink{0000-0003-3870-7215}, 
J~Veitch$^{1}$\orcidlink{0000-0002-6508-0713} and 
M~Zevin$^{3,2,4}$\orcidlink{0000-0002-0147-0835}
}

\iftoggle{checklength}{
\affil{$^1$Institute for Gravitational Research, University of Glasgow, Kelvin Building, University Ave., Glasgow, G12 8QQ, United Kingdom} 

\affil{$^2$Center for Interdisciplinary Exploration and Research in Astrophysics (CIERA), Northwestern University, 1800 Sherman Ave., Evanston, IL 60201, USA}

\affil{$^3$The Adler Planetarium, 1300 S.\ DuSable Lake Shore Drive, Chicago, IL 60605, USA} 

\affil{$^4$NSF-Simons AI Institute for the Sky (SkAI), 172 E.\ Chestnut St., Chicago, IL 60611, USA } 
}

\email{s.colloms.1@research.gla.ac.uk}

\begin{abstract}
The growing number of gravitational-wave detections from \aclp{BBH} enables increasingly precise measurements of their population properties. 
The observed population is most likely drawn from multiple formation channels. 
Population-synthesis simulations allow detailed modelling of each of these channels, and comparing population-synthesis models with the observations allows us to constrain the uncertain physics of binary black hole formation and evolution. 
However, the most detailed population-synthesis codes are computationally expensive. 
We demonstrate the use of normalising flows to emulate five different population synthesis models, reducing the computational expense, and allowing interpolation between the populations predicted for different simulation inputs. 
With the trained normalising flows, we measure the branching ratios of different formation channels and details of binary stellar evolution, using the current catalogue of gravitational-wave observations. 
\end{abstract}

\section{The challenge of population inference}

Binary stars are common, with most massive stars undergoing interactions with a companion star in their lifetime~\citep{Sana:2012px, Marchant:2023wno}.
However, there are many unknown details of binary evolution. 
Understanding the physics of formation, mass transfer, supernovae, and dynamical interactions of stellar systems is a current challenge~\citep{Duchene:2013cba, Antonini:2018auk, Bavera:2020uch, Belczynski:2021zaz,Fryer:2022lla, Burrows:2024wqv}.
We can use observations of massive stars to help constrain these uncertainties~\citep{Mapelli:2021taw, KAGRA:2021duu, Marchant:2023wno, DiCarlo:2023qey, Xing:2024hrt, Callister:2024cdx, Willcox:2025jzd}. 
\Ac{GW} detections provide constraints on the evolution of massive low-metallicity stars that are hard to observe otherwise~\citep{LIGOScientific:2016vpg,Fishbach:2021mhp}, and complement the study of of massive stars at other evolutionary stages~\citep{Eldridge:2018nop,Bavera:2021ukj,Liotine:2022vwq}. 
To extract the information encoded in observations requires carefully constructing an accurate inference framework.

To understand the astrophysics of the \ac{GW} population, we can compare \ac{GW} observations to simulations which explore astrophysical uncertainties as variable input parameters~\citep{Stevenson:2017tfq, Zevin:2017evb, Neijssel:2019irh, Bouffanais:2020qds, Zevin:2020gbd, Bavera:2020uch, Mastrogiovanni:2022ykr, Stevenson:2022djs}.
Varying the inputs to these simulations allows us to compare different astrophysical assumptions to the data. 
By inferring the most probable astrophysical assumptions using \ac{GW} data, we aim to investigate the population parameter space and the correlations between population parameters~\citep{Barrett:2017fcw,Colloms:2025hib}. 
Constraining the population parameters that describe the detailed astrophysics of binary evolution reduces the uncertainties in our understanding of the origins of merging binaries. 
However, detailed simulations of binary evolution are computationally expensive~\cite{Breivik:2025edm}.
We need to simulate many binaries per population to cover the binary parameter space, and many populations to cover the population parameter space.
Models with detailed simulations of stellar and binary physics, including mass transfer, angular-momentum transport, and supernovae, can take $\mathcal{O}(10^2)$ CPU hours for a single population of $\mathcal{O}(10^6)$ binaries~\citep{Andrews:2024saw}.

We show how normalising flows can be used to alleviate the expense of simulating many populations.  
Normalising flows are a machine-learning technique, especially suitable for emulating distributions~\citep{Papamakarios:2019fms, Kobyzev:2019ydm,Wong:2020jdt}. 
We have trained normalising flows to emulate the distribution of observable parameters for a variety of binary formation channels~\citep{Colloms:2025hib}. 
With the normalising flows, we can evaluate these distributions at any location in the training population parameter space. 
Hence, we can interpolate between the population parameters at which the populations were originally simulated. 
The normalising flows have been implemented within the \ac{AMAZE} framework~\citep{Zevin:2020gbd,Colloms:2025hib} to infer astrophysical input parameters from \ac{GW} observations.
By incorporating normalising flows, we obtain posteriors over a continuous range of population parameters, overcoming the limitation of only being able to compare data to a model where simulations were originally performed.

\section{Emulation and inference}

We train the normalising flows to learn a distribution of \ac{BBH} properties $\vec{\theta}$ given population parameters $\vec{\lambda}$ for five formation channels. 
The channels include three isolated binary populations: common envelope~\citep{1976Paczynski,vandenHeuvel1976, Dominik:2012kk,Ivanova:2012vx}, stable mass transfer~\citep{vandenHeuvel:2017pwp, Neijssel:2019irh, Gallegos-Garcia:2021hti, vanSon:2021zpk, Briel:2022cfl} and chemically homogeneous evolution~\citep{deMink:2008iu, Mandel:2015qlu, duBuisson:2020asn}, plus two stellar-cluster environments which consider dynamical interactions: globular clusters and nuclear star clusters~\citep{Heggie:1975tg, Fitchett:1983qzq, Antonini:2016gqe, Rodriguez:2019huv}. 
We explore two population parameters $\vec{\lambda}$~\citep{Zevin:2020gbd}: the natal spin of black holes born in isolation $\chib$, and the common-envelope efficiency $\alphaCE$, which only affects the distribution of the common-envelope channel. 
The observable \ac{BBH} properties $\vec{\theta}$ are chirp mass $\mathcal{M}$, mass ratio $q$, effective inspiral spin $\chi_{\mathrm{eff}}$, and redshift $z$. 
While distributions for $\vec{\theta}$ are only available at certain values of $\vec{\lambda}$ where the population-synthesis simulations were generated, the normalising flows can interpolate between these values to predict $p(\vec{\theta}\mid\vec{\lambda})$ at any $\chib$ and $\alphaCE$ within the simulated range. 
The interpolation enables us to evaluate the posterior distribution $p(\vec{\lambda}\mid\{d\})$ given our set of observational data $\{d\}$.
\begin{figure*}
    \centering
    \includegraphics[width=0.75\textwidth]{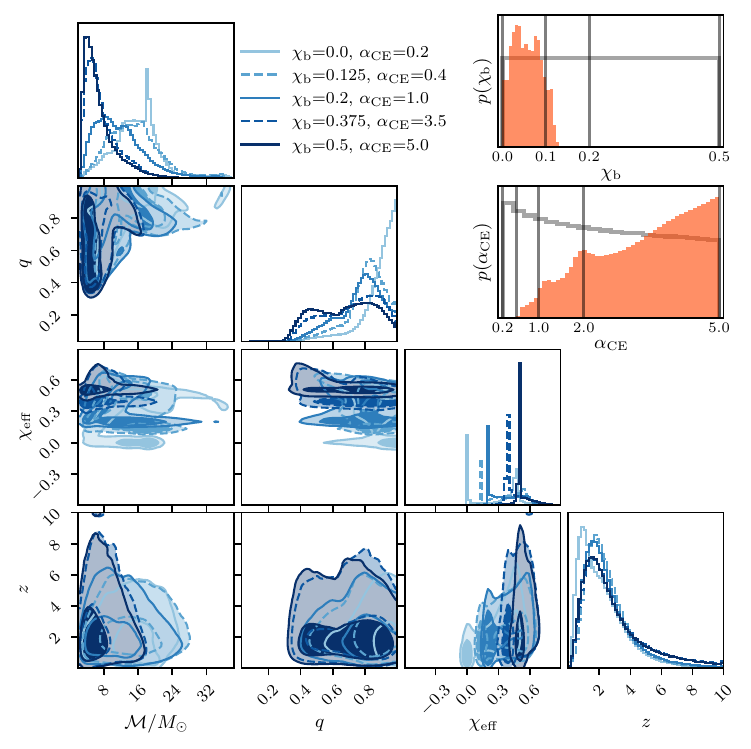}
    \caption{\emph{Bottom left:} The normalising-flow emulated distributions of chirp mass, mass ratio, effective inspiral spin and redshift for the common-envelope channel at five values of natal spin $\chib$ and common-envelope efficiency $\alphaCE$. 
    The solid lines show the emulated distribution evaluated where there is training data from population synthesis. 
    The dashed distributions are evaluated at $\{\chib, \alphaCE\}$ where there is no training data, showing the interpolation of the normalising flow. 
    \emph{Top right:} The posterior distributions for $\chib$ and $\alphaCE$ (orange) using normalising flows for continuous inference~\citep{Colloms:2025hib}, using observations from GWTC-3.0~\citep{LIGOScientific:2021djp}. 
    The $\chib$ and $\alphaCE$ values where population synthesis data used for training are shown by the grey vertical lines. 
    The prior distributions are shown in grey.}
    \label{fig:interpresults}
\end{figure*}

The normalising flow is based around a neural network which is trained to predict a target distribution~\citep{Papamakarios:2019fms,Kobyzev:2019ydm}. 
The emulation ability of the normalising flows depends on parameters of the neural network configuration, which can be tuned to control the training process~\citep{Dewancker:2016iht, zhang2019deep, koehler21a, Jiang2024}. 
To train the normalising flows, we chose these parameters to optimise the performance of the networks, found with the optimisation tool Weights and Biases \citep{wandb}. 

The trained normalising flows are better emulators of our training data than previously-used \acp{KDE}~\citep{Zevin:2020gbd} of the target distributions. 
We use the \ac{KL}~\citep{Kullback:1951zyt} divergence to measure the agreement between the target data and the emulated distributions. 
The following comparisons all quote the KL divergences respective to the target data, where differences in KL divergences indicate the relative performance of two different emulators. 
A negative value indicates that the former is a better match to the target distribution. 
The normalising flows result in a lower average \ac{KL} divergence than the \acp{KDE}: the average difference in \ac{KL} divergence being between $\KLdiffCE~\mathrm{nat}$ and $\KLdiffSMT~\mathrm{nat}$ for the channels, indicating that the normalising flow is performing better. 

Figure~\ref{fig:interpresults} (bottom left) shows the interpolation of the normalising flow between training points for the common-envelope channel in blue. 
The solid lines show the distributions at training points, with the darkest and lightest values at the two extremes of the training set.
We also plot with dashed lines the normalising-flow distribution between the simulated populations, showing how the normalising flow interpolates these distributions. 
We verify the interpolation using the common-envelope normalising flow by leaving out the training data at one $\vec{\lambda}$ point~\citep{Colloms:2025hib}. 
We find that this normalising flow reproduces the test data with a similar performance to the normalising flow trained on all the available data, with a difference in \ac{KL} divergence between the normalising flows with and without the removed data of $\KLdifftest~\mathrm{nat}$. 
The difference in \ac{KL} divergence between normalising flow with data removed and a \ac{KDE} representation of the test data is $\KLdifftestKDE~\mathrm{nat}$. 
This indicates that this normalising flow can reproduce a subset of data that was not in the original training set better than a \ac{KDE} of the training data at the removed point. 
The normalising flows are therefore trained well enough, with sufficient training data, to interpolate this set of training data. 

Once we have trained normalising flows, we can use them for hierarchical inference over a continuous range of $\chib$ and $\alphaCE$, and of the formation-channel branching fractions. 
We show the $\chib$ and $\alphaCE$ posteriors using \ac{BBH} observations from GWTC-3.0 \citep{LIGOScientific:2021djp} in Figure~\ref{fig:interpresults} (top right). 
This inference allows us to evaluate the uncertainty on our measurements of astrophysical parameters. 
We find that there is preference for low natal spins of $\chib=\chibcontall$, and high common-envelope efficiency with $\alphaCE>\alphaCEcontlowlim$ at $90\%$ credibility~\citep{Colloms:2025hib}. 
The common-envelope channel dominates the underlying population with branching fraction $\BetaCEcontall$, while the detected branching fractions are more evenly split between the five channels. 

\section{Summary}

The uncertainties of binary stellar astrophysics can be constrained by comparing observations to detailed population-synthesis simulations. 
To overcome the computational cost of running many simulations, we demonstrate the use of normalising flows to emulate population-synthesis distributions. 
The normalising flows can interpolate between simulated populations, allowing us to infer a posterior over a continuous range of astrophysical parameters. 
Our tests show that normalising flows are robust interpolators for a diverse range of population distributions, and can be used in continuous inference over multiple population parameters and multiple formation channels \citep{Colloms:2025hib}. 

We infer astrophysical parameters of the \ac{BBH} population over a continuous range with GWTC-3.0 observations~\citep{LIGOScientific:2021djp}, using normalising flows to interpolate population-synthesis distributions for multiple formation channels. 
We find preference for low natal spins and high common-envelope efficiencies, with a mix of formation channels contributing to the population of merging \acp{BBH}~\citep{Colloms:2025hib}. 
The preference for high common-envelope efficiencies indicates that more than just the orbital energy is used to eject the common-envelope to form \acp{BBH}; theoretical calculations indicate that $\alphaCE>1$ is plausible, but the maximum potential value is currently uncertain~\citep{Ivanova:2012vx, Iaconi:2019swc, Law-Smith:2020jwf,Lau:2021jpm,  Moreno:2021otq, Roepke:2022icg}. 
Our interpolated populations can be used to identify interesting regions for follow-up simulations to improve our understanding of the population parameter space most preferred by the data. 
In the future, this method can be extended to a larger range of \ac{BBH} population parameters to further explore the uncertain astrophysics of \ac{BBH}.

\iftoggle{checklength}{
\section*{Acknowledgments}
We thank Cailin Plunkett for helpful comments on the draft manuscript and Narenraju Nagarajan for useful discussions. 
SC is supported by \ac{STFC} studentship 2748218.
CPLB and JV are supported by \ac{STFC} grant ST/V005634/1. 
MZ gratefully acknowledges funding from the Brinson Foundation in support of astrophysics research at the Adler Planetarium. 
This research has made use of data from the Gravitational Wave Open Science Center, a service of the LIGO Scientific Collaboration, the Virgo Collaboration, and KAGRA. 
This material is based upon work supported by NSF's LIGO Laboratory which is a major facility fully funded by the National Science Foundation, as well as \ac{STFC} of the United Kingdom, the Max-Planck-Society (MPS), and the State of Niedersachsen/Germany for support of the construction of Advanced LIGO and construction and operation of the GEO\,600 detector. Additional support for Advanced LIGO was provided by the Australian Research Council. 
Virgo is funded, through the European Gravitational Observatory (EGO), by the French Centre National de Recherche Scientifique (CNRS), the Italian Istituto Nazionale di Fisica Nucleare (INFN) and the Dutch Nikhef, with contributions by institutions from Belgium, Germany, Greece, Hungary, Ireland, Japan, Monaco, Poland, Portugal, Spain. KAGRA is supported by Ministry of Education, Culture, Sports, Science and Technology (MEXT), Japan Society for the Promotion of Science (JSPS) in Japan; National Research Foundation (NRF) and Ministry of Science and ICT (MSIT) in Korea; Academia Sinica (AS) and National Science and Technology Council (NSTC) in Taiwan. 
This document has been assigned LIGO document number LIGO-P2500510.
}

\bibliographystyle{iopart-num}
\bibliography{AMAZE-ref}

\end{document}

%% file: result_macros_extra.tex
\newcommand{\KLdiffCE}{-0.37}
\newcommand{\KLdiffCHE}{-0.49}
\newcommand{\KLdiffGC}{-0.67}
\newcommand{\KLdiffNSC}{-0.72}
\newcommand{\KLdiffSMT}{-0.97}
\newcommand{\BFchidisc}{1.3}
\newcommand{\BFalphaCEdisc}{2.7}
\newcommand{\BFchiKDE}{0.61}
\newcommand{\BFalphaCEKDE}{2.1}
\newcommand{\alphaCEcontlowlim}{3.7}
\newcommand{\chibcontSDR}{2.5}
\newcommand{\BetaCEcontall}{0.908^{+0.045}_{-0.102}}
\newcommand{\BetaCEdiscall}{0.897^{+0.049}_{-0.097}}
\newcommand{\BetaCEKDEall}{0.913^{+0.042}_{-0.088}}
\newcommand{\BetaCHEcontall}{0.005^{+0.009}_{-0.004}}
\newcommand{\BetaCHEdiscall}{0.006^{+0.010}_{-0.004}}
\newcommand{\BetaCHEKDEall}{0.005^{+0.009}_{-0.004}}
\newcommand{\BetaGCcontall}{0.015^{+0.046}_{-0.014}}
\newcommand{\BetaGCdiscall}{0.048^{+0.061}_{-0.034}}
\newcommand{\BetaGCKDEall}{0.032^{+0.056}_{-0.029}}
\newcommand{\BetaNSCcontall}{0.013^{+0.015}_{-0.008}}
\newcommand{\BetaNSCdiscall}{0.016^{+0.018}_{-0.010}}
\newcommand{\BetaNSCKDEall}{0.015^{+0.017}_{-0.009}}
\newcommand{\BetaSMTcontall}{0.053^{+0.079}_{-0.036}}
\newcommand{\BetaSMTdiscall}{0.029^{+0.051}_{-0.024}}
\newcommand{\BetaSMTKDEall}{0.031^{+0.055}_{-0.026}}
\newcommand{\chibcontall}{0.04^{+0.04}_{-0.01}}
\newcommand{\alphaCEcontall}{4.6^{+0.4}_{-1.2}}
\newcommand{\chibdiscall}{1.00^{+0.00}_{-0.00}}
\newcommand{\alphaCEdiscall}{4.0^{+0.0}_{-0.0}}
\newcommand{\chibKDEall}{1.00^{+0.00}_{-1.00}}
\newcommand{\alphaCEKDEall}{4.0^{+0.0}_{-0.0}}
\newcommand{\KLdifftest}{-0.04}
\newcommand{\KLdifftestKDE}{-0.41}
\newcommand{\chieffKSratio}{10}
\newcommand{\highchibfrac}{14.5}
\newcommand{\upperchieffpercentile}{0.63}
\newcommand{\highchibfracparam}{0.2}
\newcommand{\skewhighchibfracparam}{1.6}
\newcommand{\BetaDetCEall}{0.22^{+0.13}_{-0.10}}
\newcommand{\BetaDetCHEall}{0.06^{+0.08}_{-0.05}}
\newcommand{\BetaDetGCall}{0.10^{+0.22}_{-0.09}}
\newcommand{\BetaDetNSCall}{0.25^{+0.16}_{-0.13}}
\newcommand{\BetaDetSMTall}{0.33^{+0.21}_{-0.21}}
\newcommand{\BetaDetPercCEall}{22^{+13}_{-10}}
\newcommand{\BetaDetPercCHEall}{6^{+8}_{-5}}
\newcommand{\BetaDetPercGCall}{10^{+22}_{-9}}
\newcommand{\BetaDetPercNSCall}{25^{+16}_{-13}}
\newcommand{\BetaDetPercSMTall}{33^{+21}_{-21}}
\newcommand{\BetaCEcont}{0.908}
\newcommand{\BetaCEcontup}{0.045}
\newcommand{\BetaCEcontlow}{0.102}
\newcommand{\BetaCEdisc}{0.897}
\newcommand{\BetaCEdiscup}{0.049}
\newcommand{\BetaCEdisclow}{0.097}
\newcommand{\BetaCEKDE}{0.913}
\newcommand{\BetaCEKDEup}{0.042}
\newcommand{\BetaCEKDElow}{0.088}
\newcommand{\BetaCHEcont}{0.005}
\newcommand{\BetaCHEcontup}{0.009}
\newcommand{\BetaCHEcontlow}{0.004}
\newcommand{\BetaCHEdisc}{0.006}
\newcommand{\BetaCHEdiscup}{0.010}
\newcommand{\BetaCHEdisclow}{0.004}
\newcommand{\BetaCHEKDE}{0.005}
\newcommand{\BetaCHEKDEup}{0.009}
\newcommand{\BetaCHEKDElow}{0.004}
\newcommand{\BetaGCcont}{0.015}
\newcommand{\BetaGCcontup}{0.046}
\newcommand{\BetaGCcontlow}{0.014}
\newcommand{\BetaGCdisc}{0.048}
\newcommand{\BetaGCdiscup}{0.061}
\newcommand{\BetaGCdisclow}{0.034}
\newcommand{\BetaGCKDE}{0.032}
\newcommand{\BetaGCKDEup}{0.056}
\newcommand{\BetaGCKDElow}{0.029}
\newcommand{\BetaNSCcont}{0.013}
\newcommand{\BetaNSCcontup}{0.015}
\newcommand{\BetaNSCcontlow}{0.008}
\newcommand{\BetaNSCdisc}{0.016}
\newcommand{\BetaNSCdiscup}{0.018}
\newcommand{\BetaNSCdisclow}{0.010}
\newcommand{\BetaNSCKDE}{0.015}
\newcommand{\BetaNSCKDEup}{0.017}
\newcommand{\BetaNSCKDElow}{0.009}
\newcommand{\BetaSMTcont}{0.053}
\newcommand{\BetaSMTcontup}{0.079}
\newcommand{\BetaSMTcontlow}{0.036}
\newcommand{\BetaSMTdisc}{0.029}
\newcommand{\BetaSMTdiscup}{0.051}
\newcommand{\BetaSMTdisclow}{0.024}
\newcommand{\BetaSMTKDE}{0.031}
\newcommand{\BetaSMTKDEup}{0.055}
\newcommand{\BetaSMTKDElow}{0.026}
\newcommand{\chibcont}{0.04}
\newcommand{\chibcontup}{0.04}
\newcommand{\chibcontlow}{0.01}
\newcommand{\alphaCEcont}{4.6}
\newcommand{\alphaCEcontup}{0.4}
\newcommand{\alphaCEcontlow}{1.2}
\newcommand{\chibdisc}{1.00}
\newcommand{\chibdiscup}{0.00}
\newcommand{\chibdisclow}{0.00}
\newcommand{\alphaCEdisc}{4.0}
\newcommand{\alphaCEdiscup}{0.0}
\newcommand{\alphaCEdisclow}{0.0}
\newcommand{\chibKDE}{1.00}
\newcommand{\chibKDEup}{0.00}
\newcommand{\chibKDElow}{1.00}
\newcommand{\alphaCEKDE}{4.0}
\newcommand{\alphaCEKDEup}{0.0}
\newcommand{\alphaCEKDElow}{0.0}